\documentclass[usenatbib,useAMS]{mn2e}

\usepackage{psfig}

\newcommand{\etal}{et~al.\null}			
\newcommand{\kmsmpc}{km~s$^{-1}$~Mpc$^{-1}$}	
\newcommand{\hst}{{\it HST\/}}			
\newcommand{\chisq}{\chi_{{}_r}^2}               

\newcommand{\aap}{A\&A}
\newcommand{\aaps}{A\&A, Suppl.\ Ser.}
\newcommand{\aj}{AJ}
\newcommand{\apj}{ApJ}
\newcommand{\apjl}{ApJL}

\newcommand{\mnras}{MNRAS}
\newcommand{\nat}{Nature}

\title[Old ellipticals at $z\simeq1.5$]{Old elliptical galaxies at
$\bmath{z\simeq1.5}$ and the Kormendy relation}

\author[Waddington et al.]{I. Waddington$^{1,2}$\thanks{Email:
I.Waddington@bristol.ac.uk}, R.~A.~Windhorst$^2$, S.~H.~Cohen$^2$,
J.~S.~Dunlop$^3$, J.~A.~Peacock$^3$, 
\newauthor R. Jimenez$^4$, R.~J.~McLure$^5$, A.~J.~Bunker$^{6,7}$,
H.~Spinrad$^7$, A.~Dey$^8$ and D.~Stern$^9$\\
$^1$ Department of Physics, University of Bristol, H. H. Wills Physics
Laboratory, Tyndall Avenue, Bristol, BS8 1TL, UK\\
$^2$ Department of Physics \& Astronomy, Arizona State University, PO
Box 871504, Tempe, AZ 85287--1504, USA\\ 
$^3$ Institute for Astronomy, University of Edinburgh, Royal
Observatory, Blackford Hill, Edinburgh EH9 3HJ, UK\\
$^4$ Department of Physics \& Astronomy, Rutgers University, 136
Frelinghuysen Road, Piscataway, NJ~08854--8019, USA\\
$^5$ Nuclear and Astrophysics Laboratory, University of Oxford, Keble
Road, Oxford, OX1 3RH, UK\\
$^6$ Institute of Astronomy, University of Cambridge, Madingley Road,
Cambridge, CB3 0HA, UK\\
$^7$ Astronomy Department, University of California at Berkeley, 601
Campbell Hall, Berkeley, CA~94720--3411, USA\\
$^8$ KPNO/NOAO, 950 North Cherry Avenue, P. O. Box 26732, Tucson,
AZ~85726, USA\\
$^9$ Jet Propulsion Laboratory, California Institute of Technology,
Mail Stop 169--327, Pasadena, CA~91109, USA
}

\date{\it Accepted 2002/07/19.}

\pagerange{\pageref{firstpage}--\pageref{lastpage}}
\pubyear{2002}

\begin{document}

\maketitle

\label{firstpage}

\begin{abstract}

Deep spectroscopy of the two millijansky radio galaxies LBDS 53W069
and LBDS 53W091 has previously shown them to have old ($\ga3$~Gyr)
stellar populations at $z\simeq1.5$.  Here we present the results of
{\it Hubble Space Telescope (HST)\/} observations with the Wide Field
and Planetary Camera 2 (WFPC2) in F814W and with the Near-Infrared
Camera and Multi-Object Spectrograph (NICMOS) in F110W.  We find that
53W069 has a de Vaucouleurs $r^{1/4}$ profile in both the F814W \&
F110W data with a mean effective radius of $0\farcs30\pm0\farcs06$
($2.7\pm0.5$~kpc).  The restframe $U-B$ colour gradient is consistent
with that of present-day ellipticals, requiring a stellar population of
super-solar ($3Z_{\sun}$) metallicity that formed on a very short
timescale at high redshift ($z>5$).  53W091 has a regular $r^{1/4}$
profile in F110W with an effective radius of $0\farcs32\pm0\farcs08$
($2.9\pm0.7$~kpc).  The F814W profile is more extended and is
consistent with the presence of a blue exponential disk that
contributes $20\pm10$\% of the flux within $r_e$.  We find a restframe
$U-B$ colour gradient which is significantly larger than that observed
in field ellipticals at $z\le1$, implying a stellar population of
mixed metallicity (1--$3Z_{\sun}$) that formed in a high-redshift
rapid burst.

We have compared these two LBDS radio galaxies with the Kormendy
relations of ten 3CR radio galaxies at $z\simeq0.8$ and a sample of
cluster ellipticals at $z\sim0.4$.  The LBDS galaxies follow the
Kormendy relation for the more radio-luminous 3CR galaxies, assuming
passive evolution of their stellar populations, although they are
smaller than the 3CR galaxies whose mean effective radius is 12~kpc.
Their sizes and radio luminosities are consistent with scaling
relations applied to the 3CR galaxies, in which both radio power and
effective radius scale with galaxy mass.  Compared with the sample of
cluster ellipticals, 53W069 \& 53W091 lie well within the scatter of
the Kormendy relation.  We conclude that the hosts of these
millijansky radio sources at $z\simeq 1.5$ are passively-evolving
elliptical galaxies that will evolve into ordinary $L^*$ ellipticals
by the present day.

\end{abstract}

\begin{keywords}
galaxies: active --- galaxies: elliptical and lenticular, cD
--- galaxies: evolution --- galaxies: individual (LBDS 53W069, LBDS
53W091)
\end{keywords}

\vspace*{10mm}

\section{Introduction}

How did elliptical galaxies form and evolve?  There are two mechanisms
that have been proposed.  First is the monolithic collapse model, in
which elliptical galaxies form at high redshifts in an intense burst
of star formation \citep{Eggen62}.  Essentially all the final mass of
the galaxy is already present within its potential well at the time of
formation.  The second mechanism is the merger model, in which
ellipticals form at relatively lower redshifts from the merger of many
smaller galaxies \citep[e.g.,][]{Kauffmann98}.  In this model, the
mass of the galaxy increases with time as more and more smaller
galaxies are canibalized by the forming elliptical.

Recent observations suggest that both mechanisms may have a role to
play.  For example, \citet{Ellis97} found that spheroidal galaxies in
clusters at $z\simeq0.5$ required a formation redshift (by which we
mean the redshift at which the dominant stellar population formed) of
$z_f\ge3$.  In contrast, \citet{Zepf97} found that the small number of
very red galaxies in deep optical/infrared surveys suggested either
that ellipticals form at moderate redshifts (and are enshrouded by
dust), or that they assemble through the merging of smaller galaxies.
Similarly \citet{Pascarelle96} identified a group of Lyman-$\alpha$
emitters at $z=2.39$ that they suggested may subsequently merge at a
lower redshift into one or more luminous galaxies.  \citet{Treu99}
concluded from their study of NICMOS parallel fields that a
significant fraction (10--66\%) of the elliptical galaxy population
formed at $z_f\ge3$, but the rest may have been formed (or at least
assembled) at lower redshifts.

One important method of studying high-redshift ($z\ga1$) ellipticals
is via radio selection.  At low redshifts, luminous radio sources are
almost exclusively hosted by giant elliptical galaxies containing old
stellar populations \citep*[e.g., ][]{Kron85,Taylor96,Nolan00a}.  To
the extent that radio galaxies at high redshifts are also hosted by
giant ellipticals, they can be used to study the evolution of the
elliptical galaxy population.  \citet{Lilly84} obtained infrared $K$
magnitudes for a subsample of the 3CR radio survey, from which they
constructed a $K$-band Hubble diagram. They concluded from this
$K$--$z$ relation that luminous radio galaxies at $z\sim1$ are giant
ellipticals with passively-evolving stellar populations.
\citet{Lilly85} and \citet{Dunlop90} found that the $K$--$z$ relation
for less-powerful radio galaxies led to a similar conclusion.

At redshifts $z\ga0.6$, \citet{Eales97} found that the 6C radio
galaxies, with radio luminosities a factor of $\sim5$ lower than the
3CR, were also on average 0.6~mag fainter than 3CR in the $K$-band.
\citet{Roche98} investigated the $K^\prime$-band morphologies of ten
of these 6C sources at $z\sim1.1$, showing that seven of them were
normal ellipticals and the other three were ongoing or recent mergers.
The radii of the 6C galaxies were significantly smaller than those of
the 3CR sources at similar redshifts \citep{Best98,Zirm99,McLure00},
indicating that their fainter $K$ magnitudes were due to their smaller
size, not solely due to the difference in power of their AGN.

The Leiden-Berkeley Deep Survey \citep[LBDS;
][]{Windhorst84,Waddington00}, with a flux density limit of 1~mJy at
1.4~GHz, is sensitive to radio sources a factor of $\sim200$ fainter
than the 6C survey, thus probing lower radio luminosities and higher
redshifts.  \citet*{Windhorst98} showed that the radio galaxy LBDS
53W002 has a weak AGN and a dominant $r^{1/4}$ profile with the
colours of a $\sim 3\times 10^8$~year young stellar population at
$z=2.39$, suggesting that these weak radio sources could have formed
rather mature ellipticals at high redshifts.  Two more of these
sources have proved particularly useful for the study of galaxy
evolution, due to the deep spectra that we obtained with the Keck
telescope.  LBDS 53W091 has a redshift of 1.552, and its restframe UV
spectrum is best modeled by a stellar population $\ge3.5$~Gyr old
\citep{Dunlop96,Spinrad97}.  LBDS 53W069 is at $z=1.432$ and has a
best-fitting stellar population of age $\ge4$~Gyr
\citep{Dey97,Dunlop99b}.  \citet{Spinrad97} investigated the age
determination in detail and showed how the minimum age of the universe
at $z\simeq1.5$, as required by the age of 53W091, placed limits on
the allowed values of the cosmological parameters ($H_0$, $\Omega_{\rm
M}$, $\Omega_\Lambda$).  \citet{Stockton95} similarly showed that the
age inferred from the Keck spectrum of the radio galaxy 3C65
($\sim4$~Gyr at $z=1.175$) required a formation redshift of $z_f\ge5$.

Several authors have questioned the reliability of these age
estimates, for example \citet{Bruzual99} and \citet{Yi00} derive ages
of 1.5--2~Gyr for 53W091.  However, \citet{Dunlop99} argued that such
young ages are only deduced if the near-infrared photometry is
included in the model fitting; if the fitting is confined to the
spectroscopic data then a variety of stellar population sythesis codes
consistently produce ages of $\ge2.5$~Gyr (see also Nolan \etal\
2001b\nocite{Nolan00b}).  We will not discuss this age controversy
further here, but simply note that nothing in the current paper
depends crucially on the precise ages of these sources.

In this paper we will investigate the morphologies of these two radio
galaxies with the {\it Hubble Space Telescope's (HST)} Wide Field and
Planetary Camera 2 (WFPC2) and Near-Infrared Camera and Multi-Object
Spectrograph (NICMOS).  At $z\simeq1.5$, the 4000-\AA\ break is
straddled by the F814W and F110W filters, and thus the emission
observed through these two filters is dominated by the young and old
stellar populations respectively.  In section 2, we describe the
observations and discuss the processing steps that were applied to the
data.  In section 3, we investigate the surface brightness profiles
and colour gradients of the two sources.  The location of the galaxies
on the Kormendy relation is the topic of section 4 and finally, in
section 5, we comment on the apparent relations between radio
luminosity, black-hole mass and effective radius of these sources.
Our results are summarized in section 6.

We use AB magnitudes unless otherwise noted, and denote magnitudes in
the three \hst\ filters by $I_{814}$ for F814W (WFPC2), $J_{110}$ for
F110W (NICMOS) and $H_{160}$ for F160W (NICMOS).  We assume a flat
cosmology with $H_0=65$~\kmsmpc, $\Omega_{\rm M}=0.3$,
$\Omega_\Lambda=0.7$.

\section{Observations and data reduction}

\begin{figure*}
\begin{minipage}{150mm}
\psfig{file=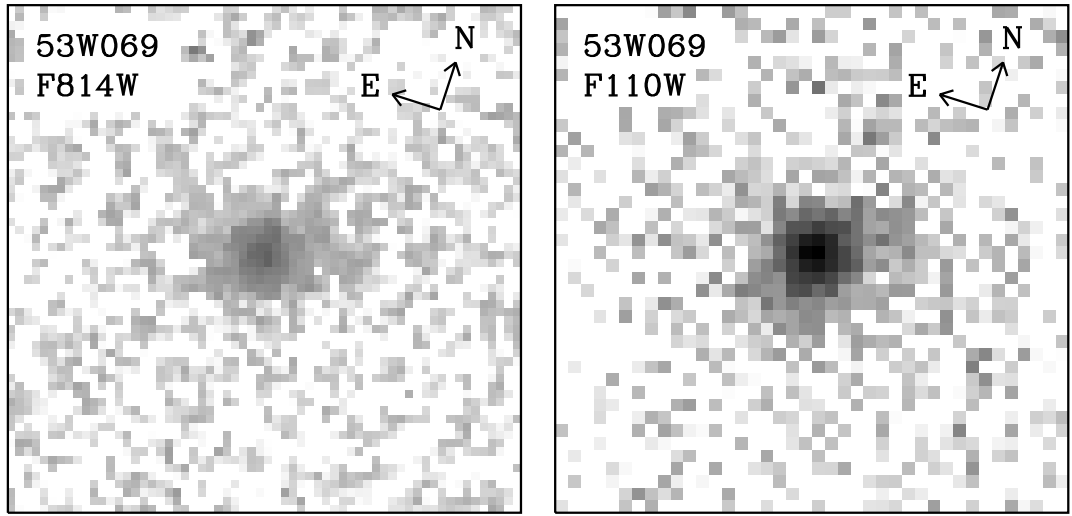,width=150mm}
\caption{WFPC2 F814W and NICMOS F110W images of 53W069.  The
logarithmic grey-scale has been chosen so that both images have the
same range in flux density ($F_\lambda$).  Each image is
$3\times3$~arcsec$^2$.  The F814W image has been rotated in order to
display it at the same orientation as the F110W
image.\label{w69images}}
\end{minipage}
\end{figure*}

In 1997 August and 1997 October, we observed 53W069 and 53W091 with
WFPC2 and NICMOS on the \hst.  53W069 was observed for 2 orbits (a
total integration of 5100~s) in F814W with WFPC2, and for 3 orbits
(7680~s) in F110W with NICMOS camera 2.  53W091 was observed for 2
orbits (5300~s) in F814W with WFPC2, and for 2 orbits (5120~s) in
F110W with NICMOS camera 2.  Each orbit devoted to WFPC2 imaging was
split into two exposures, and consecutive orbits were offset by
0\farcs35.  In each orbit devoted to NICMOS imaging, five 512~s
exposures were taken using a spiral dither pattern with 1\farcs91
offsets.  In 1998 August, we observed 53W091 and its companions for 10
orbits (25$\,$600~s) in F160W with NICMOS camera 2.  We independently
analyse those data and describe the results in a complementary paper
\citep{Bunker00}.

\begin{figure*}
\begin{minipage}{150mm}
\psfig{file=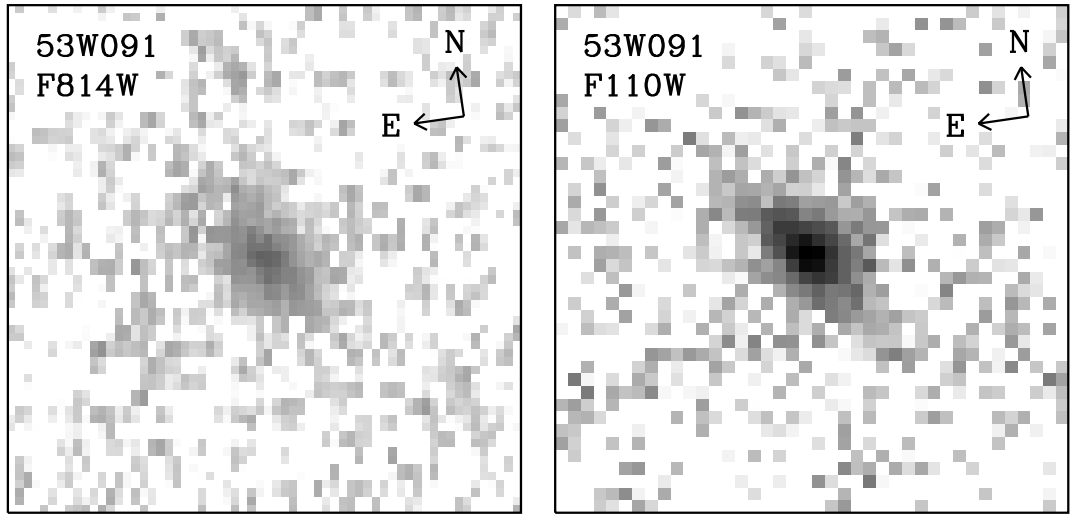,width=150mm}
\caption{WFPC2 F814W and NICMOS F110W images of 53W091.  The
logarithmic grey-scale has been chosen so that both images have the
same range in flux density ($F_\lambda$).  Each image is
$3\times3$~arcsec$^2$.  The F814W image has been rotated in order to
display it at the same orientation as the F110W image, resulting in
the apparent blurring of the drizzled data.\label{w91images}}
\end{minipage}
\end{figure*}

The WFPC2 exposures were reduced using the standard pipeline
processing \citep{Voit97}.  For each target, the four dithered
exposures were combined into a single image using the drizzling
technique \citep{Hook97,Fruchter97}.  A pixel size of 0\farcs05,
one-half of the original pixel scale, was used, with a drop size of
0.6 (the parameter PIXFRACT).  Offsets were calculated from the
cross-correlation of $\sim 10$ bright sources in each exposure; no
rotation was required.  The drizzled images are shown in
Figs \ref{w69images} \& \ref{w91images}.

We reduced the NICMOS images using a combination of processing
techniques from {\sc stsdas} \citep{Dickinson99} and our own methods.
The raw images were partially processed with the standard pipeline
task {\sc calnica}, using the best reference files available from the
STScI.  We tried using the latest temperature-dependent darks in order
to remove the shading, however we obtained better results with the
standard dark reference files and used these instead.  Non-linearities
in the bias level of each image quadrant were removed with {\sc
biaseq}.  Bias jumps were not corrected for in the data, as only one
image was significantly affected and {\sc biaseq} was not successful
in removing the jumps.  Cosmic rays were then removed by continued
processing with {\sc calnica}.  The variable bias level from quadrant
to quadrant, or ``pedestal'', was corrected using the {\sc pedsky}
task, which also subtracted the sky background from the images.  Then
the data were flattened using {\sc calnica}.

\begin{table}
\caption{Photometry}
\label{photom}
\begin{tabular}{@{}lccc} \hline
Source & \multicolumn{2}{c}{Total AB magnitude} & Colour$^*$ \\
       & $I_{814}$       & $J_{110}$       & $I_{814}-J_{110}$ \\ \hline
53W069 & $24.19\pm 0.03$ & $22.45\pm 0.02$ & $1.72\pm 0.05$ \\
53W091 & $24.14\pm 0.03$ & $22.56\pm 0.03$ & $1.57\pm 0.05$ \\ \hline
\end{tabular}

\medskip
$^*$The colours are defined in a circular aperture of 2\arcsec\ diameter.
\end{table}

We applied a further correction to the data, adapted from standard
ground-based infrared imaging methods.  Since the NICMOS camera
consists of four physically separate sub-arrays, we divided each
partially-reduced image into its four separate quadrants.  For each
target, all the exposures of each quadrant were stacked and a median
image calculated.  Given that the exposures were dithered, this
produced a map of the residual instrumental features (devoid of any
astronomical objects) that were not removed by the flat-field
reference file.  These ``super-sky'' images for each quadrant were
scaled to the mean of all four quadrants in order to preserve the
quadrant-to-quadrant photometric accuracy.  We then tried correcting
each exposure by either (i) dividing by this super-sky after scaling
it to have a mean of unity, or (ii) subtracting the super-sky from
each exposure.  For each object, we chose the method that produced the
best image (i.e.\ the lowest rms noise).  For the F110W images of both
targets, {\it subtracting\/} the super-sky was most successful,
suggesting that the instrumental features left after {\sc calnica}
processing were most likely due to incomplete dark subtraction.  In
particular, this removed the residual amplifier glow from the corners,
while making a negligible difference to the center of the image, where
the target was located (thus photometric accuracy had not been
compromised by residual dark current beneath the target).  Finally,
all the exposures were combined, on a quadrant by quadrant basis, to
produce a mosaic for each object (Figs \ref{w69images} \&
\ref{w91images}).  The data were calibrated using the most recent
photometric calibration parameters from the STScI.

The image detection program {\sc sextractor} \citep{Bertin96} was used
to calculate both aperture and total (``mag\_auto'') AB magnitudes for
the two radio galaxies in each of the filters.  The results are given
in Table~\ref{photom}.  Although the total magnitudes give the best
measurement of the flux detected in each individual image, the colours
of the sources are properly defined in fixed apertures, thus we give
the colours in Table~\ref{photom} for 2\arcsec\ diameter circular
apertures.  We note that integration of the best-fitting model surface
brightness profiles (see section 3.1 below) gives magnitudes
consistent with these measurements.

\section{Morphologies}

\subsection{Surface brightness profiles}

\begin{figure}
\psfig{file=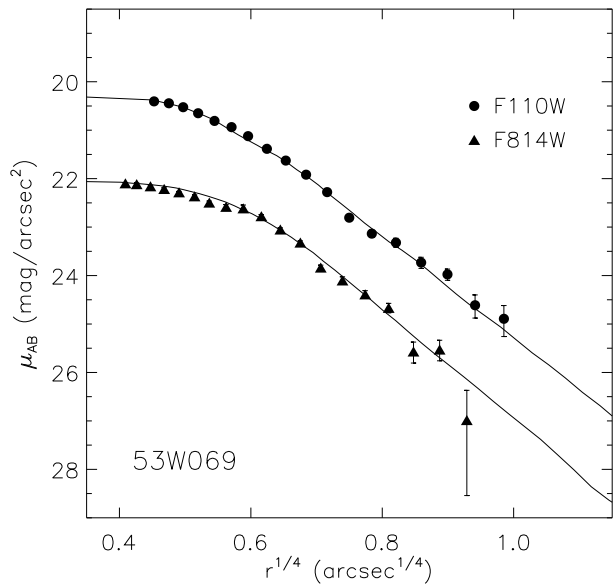,width=85mm}
\caption{Surface brightness profiles along the major axis (points) and
best-fitting models (lines) for 53W069.  Both profiles follow a de
Vaucouleurs $r^{1/4}$ law, with effective radii of $r_e=0\farcs33$ in
F110W and $r_e=0\farcs28$ in F814W.\label{w69profiles}}
\end{figure}

\begin{figure}
\psfig{file=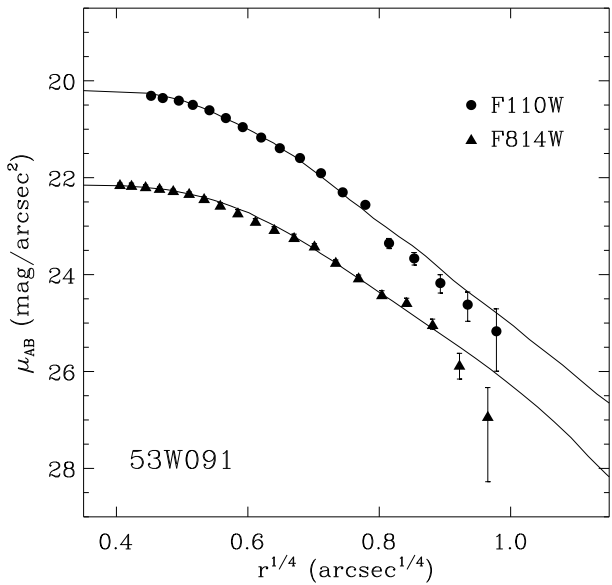,width=85mm}
\caption{Surface brightness profiles along the major axis (points) and
best-fitting models (lines) for 53W091.  The F110W profile follows a
de Vaucouleurs $r^{1/4}$ law, with effective radius of
$r_e=0\farcs32$.  The F814W profile is the sum of an $r^{1/4}$ profile
with effective radius $r_e=0\farcs32$ plus an exponential with
scale-length $r_0=0\farcs5$ contributing 20\% of the flux within
$r_e$.\label{w91profiles}}
\end{figure}

Surface brightness profiles were extracted for the two radio galaxies
by fitting elliptical isophotes to the sky-subtracted images (Figs
\ref{w69profiles} \& \ref{w91profiles}) with {\sc iraf}'s {\sc
ellipse} routine.  We fixed the centre of each isophote and allowed
the position angle and axial ratio to vary freely.  Model surface
brightness profiles were extracted in the same way from
two-dimensional exponential and de Vaucouleurs $r^{1/4}$ models
convolved with a model point-spread function (PSF) from Tiny Tim
\citep{Krist99}.  The models were subsampled by a factor of 1, 2 or 5,
convolved with a subsampled PSF and then resampled to the (drizzled)
WFPC2 and NICMOS pixel size, before being extracted to a
one-dimensional profile.  The models were fitted to the data for
$r>1$~pixel, corresponding to 0\farcs05 for WFPC2 (drizzled) and
0\farcs075 for NICMOS -- at smaller radii, the fit was dependent on
the degree of subsampling of the PSF.

The best-fitting models for 53W069 are de Vaucouleurs $r^{1/4}$
profiles in both the F814W \& F110W filters (Fig.\ \ref{w69profiles}).
In F814W we find an effective radius of $r_e=0\farcs28\pm0\farcs08$,
and in F110W the profile is very similar with
$r_e=0\farcs33\pm0\farcs09$.  The reduced chi-square for these models
is $\chisq \simeq 2$, compared with a value of $\chisq \simeq 10$ for
the best-fitting exponential models.  The weighted mean of the F814W
\& F110W effective radii is $\bar{r_e}=0\farcs30\pm0\farcs06$ or
$2.7\pm0.5$~kpc.  Morphologically, 53W069 is thus a relaxed elliptical
galaxy, fully consistent with its spectroscopic identification as an
old, passively-evolving elliptical.

For 53W091, the best-fitting model in F110W is a de Vaucouleurs
$r^{1/4}$ profile with $r_e=0\farcs32\pm0\farcs08$ (Fig.\
\ref{w91profiles}).  This model has a reduced chi-square value of
$\chisq = 1.8$ compared with a value of $\chisq = 7.3$ for the best
exponential model.  In \citet{Bunker00} we show that the F160W profile
similarly follows a de Vaucouleurs law with
$r_e=0\farcs32\pm0\farcs03$.  For the F814W profile, the best-fitting
single component model is an $r^{1/4}$ profile with
$r_e=0\farcs6\pm0\farcs2$, for which $\chisq = 3.3$ (compared with
$\chisq = 18$ for the best exponential model).  Given the large
difference between the effective radius of this best-fitting model and
that of the F110W \& F160W models, we also fitted the F814W profile
with a two-component model.  We used the best-fitting F110W/F160W de
Vaucouleurs model with $r_e=0\farcs32$ and added an exponential disk,
varying both the scale-length of the exponential and the relative flux
of the two components.  The best-fitting composite model (with $\chisq
= 2.6$) has a disk with scale-length $r_0=0\farcs5\pm0\farcs2$
($4\pm2$~kpc), that contributes 20$\pm$10\% of the flux within the
half-light radius (Fig.~\ref{w91profiles}).  At $z=1.552$, the F814W
filter measures the galaxy's restframe emission below the 4000-\AA\
break, and is thus sensitive to bluer stars than F110W.  Such a
population will make an even less significant contribution to the flux
at longer wavelengths, thus it is quite consistent that we do not see
a disk component in the F110W data.

\subsection{Colour gradients}

The stellar population of an elliptical galaxy is not uniform, but
varies with radius.  This is seen as a gradient in the colours of the
galaxy, which become bluer with increasing radius
\citep{Franx89,Peletier90,Tamura00}.  If 53W069 and 53W091 are the
passively-evolving precursors to present-day ellipticals, we should
expect to see such characteristic colour gradients.  \citet{Peletier90}
show that the $U-B$ colour gradient of nearby ellipticals is 
$-0.1$~mag~arcsec$^{-2}$ per dex in radius.  The $I_{814}$ and
$J_{110}$ bands of our radio galaxies at $z\simeq 1.5$ approximately
correspond to rest-frame $U$ and $B$, thus we should expect to see a
gradient of the same order as \citet{Peletier90}.

\begin{figure}
\psfig{file=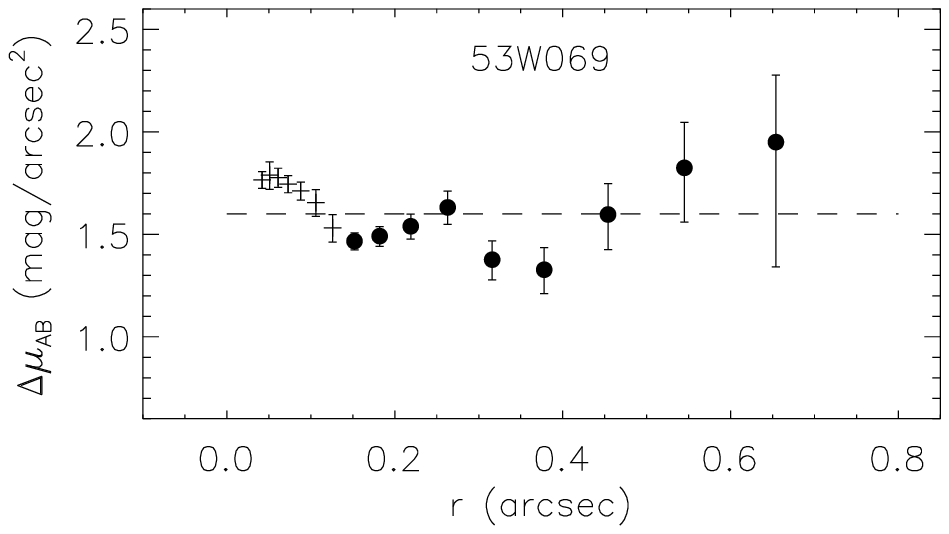,width=85mm}
\psfig{file=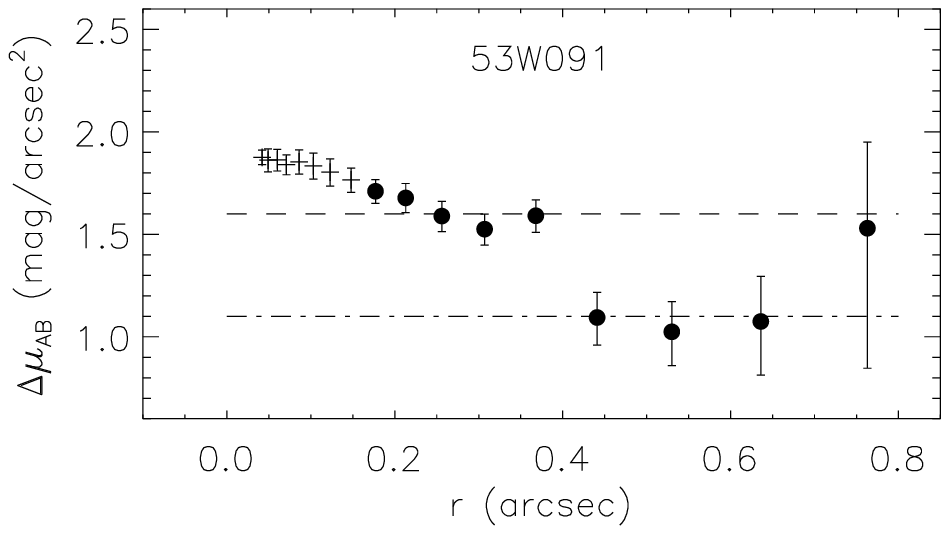,width=85mm}
\caption{$I_{814}-J_{110}$ colour gradients for 53W069 \& 53W091.
Crosses denote data within the central two pixels for which pixelation
effects and differences in the WFPC2 and NICMOS PSFs contribute as
much as 0.1--0.3~mag to the colour gradient.  The horizontal lines are
the colours predicted by a high-redshift burst of star formation, for
two different metallicities: solar (dot-dash line) and $3Z_{\sun}$
(dashed line).\label{colours}}
\end{figure}

In Fig.\ \ref{colours}, we plot the colours of the two radio galaxies
as a function of radius.  Small differences between the (drizzled)
WFPC2 and NICMOS PSFs dominate the colour gradients within the central
1--2 pixels.  Specifically, the radii containing (50\%, 90\%) of the
encircled energy are (0\farcs08, 0\farcs37) and (0\farcs08, 0\farcs48)
for the WFPC2 $F814W$ and NICMOS $F110W$ PSF models respectively.
Examination of our PSF-convolved galaxy models suggests that these
differences result in artificially redder colours within the central
pixels, contributing as much as 0.1--0.3~mag to the colour gradient at
$r<0\farcs15$ (crosses in Fig.\ \ref{colours}).  At larger radii this
effect is less than 0.1~mag, so we restrict our discussion to
$r>0\farcs15$.  Given the limited quality of the data and the absence
of PSF stars in the NICMOS images, we decided there was little to be
gained from attempting detailed PSF matching to better than 10\%.

It can be seen in Fig.\ \ref{colours} that the errors in the
$I_{814}-J_{110}$ colours of 53W069 are too large to measure a colour
change of $-0.1$~mag~arcsec$^{-2}$ across the extent of the galaxy
(less than 1~dex in radius).  Thus the colours of 53W069 do not
preclude a gradient comparable to that of present-day ellipticals, but
they are also consistent with there being no colour gradient.
\citet[][1998]{Windhorst94} similarly found little or no colour
gradients in several other LBDS radio galaxies and field galaxies at
$0.3<z<2.4$ observed with \hst.  In particular, any colour gradient in
the elliptical radio galaxy 53W002 at $z=2.39$ must be smaller than
$-0.3$~mag across the 1\arcsec\ (9~kpc) image \citep{Windhorst98}.
The (restframe) $U-B$ gradients of early-type cluster and field
galaxies at $z\simeq0.4$--1 have been found to typically span $-$0.2
to 0.1~mag per dex in radius \citep{Saglia00,Hinkley01}, again
consistent with 53W069.

In contrast, there is a significant gradient in the $I_{814}-J_{110}$
colour of 53W091.  This gradient is in the same sense as that seen in
nearby ellipticals (i.e., the galaxy is redder in the center and
becomes more blue at larger radii), but it is much greater in
magnitude.  The best fit to the colours gives a gradient of
$-1.0\pm0.2$~mag~arcsec$^{-2}$ per dex in radius.  This clearly
indicates the presence of an extended blue component in the galaxy's
morphology, and is consistent with the results of our model fitting
above.  There appears to be a ``jump'' in the colour gradient at
$r\simeq0\farcs4$ (see also the F110W profile), but this is not
significant given the errors, and the data are consistent with a
constant gradient out to $r\simeq0\farcs6$.  We note that the position
angle (PA) of the major axis of the optical/infrared emission ($\sim
45$~degrees) is perpendicular to the PA of the radio lobes
\citep[131~degrees;][]{Spinrad97}, so it is unlikely that this
extended blue emission is due to the radio--optical alignment effect.
We emphasize that the difference in WFPC2 and NICMOS PSFs is unlikely
to have produced this gradient because: (i) it is not seen in the data
for 53W069; and (ii) the profile fitting above {\it includes\/} a
convolution with the PSFs, and clearly shows that the galaxy is more
extended (i.e.\ has a larger scale-length) in the F814W data than in
the F110W data.

We compared the observed colour gradients of 53W069 and 53W091 with
the predictions of single collapse multi-zone models of elliptical
galaxies \citep*{Menanteau01}.  We used an IMF with lower and upper
cutoffs of 0.1~M${}_{\sun}$ and 120~M${}_{\sun}$ respectively, and a
range of formation redshifts.  None of the models could reproduce the
observed gradients, demonstrating that the very red cores found in
these two galaxies are inconsistent with an extended period of star
formation in their centres.  We
then modified the models to force the star formation to stop after
0.05~Gyr.  The resulting colours for two different metallicities,
solar ($Z_{\sun}$) and $3Z_{\sun}$ are shown in Fig.\ \ref{colours}.
We find that the models fit the observed gradients if star formation
took place on a short time scale in both galaxies and if they were
formed at redshift $z>5$.  53W069 is best fit by a population that
has, on average, supersolar metallicity ($3Z_{\sun}$).  53W091 has a
core that is metal-rich ($3Z_{\sun}$) and at larger radii
($r>0\farcs4$) is better fit by a solar metallicity population.  

These results are in excellent agreement with those of \citet{Nolan02}
who find that the {\it UV spectra\/} of these galaxies are similarly
best-fit with populations of solar (53W091) and supersolar
($2Z_{\sun}$, 53W069) average metallicity.  Studies of $z\simeq0.4$--1
ellipticals similarly indicate that the colour gradients of early-type
galaxies are due to metallicity gradients, falling from $\sim 2.5
Z_{\sun}$ at the centre to solar in the outer regions
\citep{Hinkley01}, although the observed gradients are not as steep as
that in 53W091.

As discussed above, the best fit to the F814W profile of 53W091 is
obtained by adding a disk-like exponential component to the
best-fitting de Vaucouleurs profile measured in the F110W and F160W
bands.  \citet{Dunlop99} showed that the difference between the
restframe UV spectrum of 53W091 and that of 53W069 can be fitted by a
blue stellar (F0V) spectrum, contributing 10--20\% of the UV flux.
Similarly, we have shown that $\sim 20$\% of the restframe $U$-band
flux within the half-light radius ($r_e=0\farcs32$) could originate in
a disk component.  With a 1\arcsec\ slit and seeing of
0\farcs8--1\arcsec, the spectroscopy of 53W091 was dominated by the
flux within $r_e$, and we find consistency between the fraction of
excess blue light in the spectroscopy and the disk component in the
imaging.  This extended blue component is also consistent with having
lower metallicity than the core of the galaxy.

Combining these results leads us to conclude that 53W091 is either:
(i) an elliptical galaxy with an old, metal-rich, passively-evolving
stellar population (just like 53W069), surrounded by a faint disk of
younger or lower-metallicity stars; or possibly (ii) an elliptical
galaxy with an old, passively-evolving stellar population whose
metallicity decreases with radius.  We note that the age of the
dominant red stellar population in 53W091 would have been somewhat
underestimated due to the presence of a younger stellar component in
the disk.  \citet{Menanteau01a} find that about half of field
ellipticals at $z\sim1$ must be undergoing recent episodes of star
formation, due to their irregular internal colours and blue cores.
The two radio galaxies 53W069 \& 53W091, in contrast, belong to the
other 50 per cent of field ellipticals that have evolved passively
since very high redshifts.

\section{The Kormendy relation}

A key characteristic of dynamically relaxed elliptical galaxies is the
existence of the Fundamental Plane \citep{Djorgovski87,Dressler87} --
a scaling relation between the luminosity $L$, effective radius $r_e$
and velocity dispersion $\sigma$ of ellipticals: $L\propto r_e^a
\sigma^b$.  Given the difficulty of measuring velocity dispersions,
particularly at high redshift, a more useful relation is the
projection of the Fundamental Plane onto the effective radius--surface
brightness ($\mu_e$) plane, or the Kormendy relation
\citep{Kormendy77}.  Cluster ellipticals \citep{Barger98,Ziegler99},
field ellipticals \citep{Schade96} and radio galaxies \citep{McLure00}
have all been shown to follow the Kormendy relation out to $z\simeq
0.5$--1.0, assuming passive evolution of their stellar populations.
Here we consider if our two high-redshift LBDS radio galaxies are
consistent with these results.  At $z\simeq1.5$ the F110W filter
approximately samples the restframe $B$-band.  We thus compare the
F110W data on 53W069 and 53W091 (Table \ref{sbdata}) with two samples
of elliptical galaxies observed with \hst\ in restframe $B$.

\begin{table}
\caption{Data for the F110W (restframe $B$-band) Kormendy relation.
$\mu_e$ is the surface brightness at the effective radius ($r_e$), and
$\left<\mu_e\right>$ is the mean surface brightness within $r_e$.}
\label{sbdata}
\begin{tabular}{@{}lccc} \hline
Source & $r_e$         & $\mu_e$             & $\left<\mu_e\right>$ \\
       & (arcsec)      & (mag arcsec$^{-2}$) & (mag arcsec$^{-2}$) \\ \hline
53W069 & $0.33\pm0.08$ & $22.6\pm0.1$        & $21.7\pm0.1$ \\
53W091 & $0.32\pm0.08$ & $22.4\pm0.1$        & $21.5\pm0.1$ \\ \hline
\end{tabular}
\end{table}

\subsection{Comparison with 3CR radio galaxies at $\bmath{z\simeq0.8}$}

\begin{figure}
\psfig{file=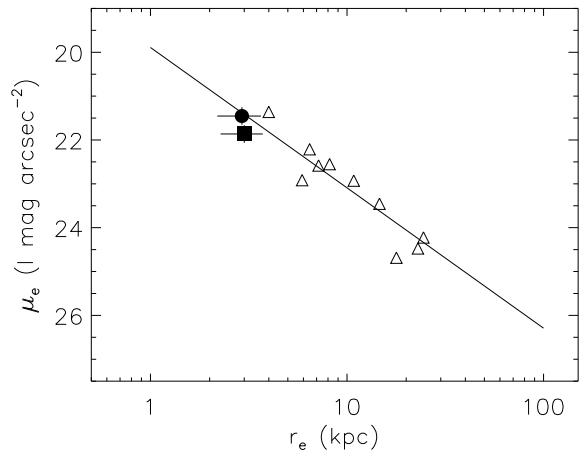,width=85mm}
\caption{Effective surface brightnesses of 53W069 \& 53W091 compared
with the Kormendy relation of a sample of 3CR radio galaxies at
$z\simeq0.8$.  53W069 (square) \& 53W091 (circle) have been corrected
for: (i) surface brightness dimming, (ii) bandpass shift (observed
F110W to $F814W$), and (iii) passive evolution, between $z\simeq
1.5$ and $z=0.8$.  The 3CR surface brightness data (triangles) have
been corrected ($\la 0.1$~mag) to a common redshift of 0.8.  The solid
line is the best-fitting Kormendy relation from \citet{McLure00},
renormalized to a redshift of 0.8 and our cosmology.\label{kormendy}}
\end{figure}

In Fig.\ \ref{kormendy}, we plot the surface brightness at the
effective radius $\mu_e$ against the effective radius $r_e$ for the
two LBDS radio galaxies and a sample of 3CR radio galaxies at
$z\simeq0.8$ \citep{McLure00}.  The F110W and F814W filters both
sample the rest-frame $B$-band at $z=1.5$ and $z=0.8$ respectively.
We have thus corrected the $J_{110}$-band surface brightness of 53W069
\& 53W091 to their equivalent $I_{814}$-band values at $z=0.8$,
minimizing any error introduced by the possible wavelength dependence
of $r_e$.  We made three corrections derived using the spectral
sythesis models of both \citet{Fioc97} and the 1996 revision of
\citet{Bruzual93}.  We used their elliptical and burst models
respectively, and refer the reader to those papers for technical
details such as IMFs.  The F110W filter is significantly wider than
F814W, thus requiring a K-correction of $+0.6$~mag to the F110W data.
Second, passive evolution of $+0.7$~mag from 3~Gyr ($z=1.5$) to 6~Gyr
($z=0.8$) was derived using the burst/elliptical models.  We set the
age of the galaxies to be 3~Gyr at $z=1.5$ and then evolved them in
our chosen cosmology until $z=0.8$.  Third, the transformation from AB
magnitudes to Vega magnitudes was $-0.7$~mag.  All the sources, LBDS
\& 3CR, were corrected for the $(1+z)^4$ surface brightness dimming,
placing them all at a common redshift of 0.8 (the 3CR sample has a
narrow redshift range of $\pm 0.07$, and correspondingly small surface
brightness corrections of $\la 0.1$~mag).  We tested the reliability
of the corrections by using the two different sets of models
(Bruzual/Charlot and Fioc/Rocca-Volmerange), and by varying the age of
the LBDS galaxies at $z=1.5$ by $\pm$1~Gyr, corresponding to the range
of spectroscopic ages derived by ourselves \citep[e.g.,][]{Dunlop96}
and others \citep[e.g.,][]{Bruzual99}.  The combined uncertainty in
K-corrections and evolution was $\pm0.2$--0.3~mag.

We have recalculated the normalization of the 3CR Kormendy relation
for our cosmology, fixing the (cosmology-independent) slope at 3.2
\citep{McLure00}, and plot the fit in Fig.\ \ref{kormendy} (solid
line).  It is seen that 53W069 and 53W091 both lie on the 3CR Kormendy
relation.  The effective radii of the two LBDS galaxies
($r_e\simeq3$~kpc) are much smaller than the mean effective radius of
the 3CR sources ($\bar{r_e}=12$~kpc), but they are comparable with the
size of the smallest 3CR galaxy in this sample (3C340 with
$r_e=4.0$~kpc).  \citet{McLure00} demonstrated that these 3CR galaxies
follow a Kormendy relation that is indistinguishable from that of AGN
hosts at $z\simeq 0.2$, assuming passive evolution of their stellar
populations.  Here we can extend that result to higher redshifts,
concluding that there is no evidence for any significant dynamical
evolution of the elliptical hosts of radio sources between $z\simeq
1.5$ and 0.2.

\subsection{Comparison with cluster ellipticals at $\bmath{z\simeq0.4}$}

\begin{figure}
\psfig{file=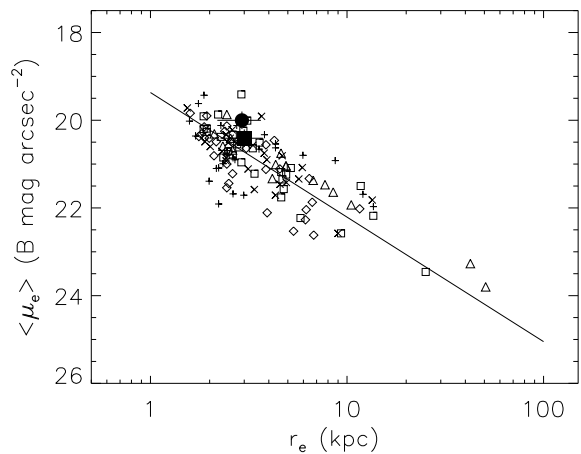,width=85mm}
\caption{Mean surface brightnesses of 53W069 \& 53W091 compared with
the Kormendy relation of a sample of cluster ellipticals at $z\sim
0.4$.  53W069 (solid square) \& 53W091 (solid circle) have been
corrected for: (i) surface brightness dimming, (ii) bandpass shift
(observed F110W to $B$), and (iii) passive evolution, between
$z\simeq 1.5$ and $z=0.0$.  The cluster ellipticals have been taken
from \citet{Ziegler99}, and have similarly been transformed to zero
redshift.  The clusters are: Abell 370 at $z=0.375$ (triangles);
Cl~1447$+$26 at $z=0.389$ (diamonds); Cl~0939$+$47 at $z=0.407$ (plus
symbols); Cl~0303$+$17 at $z=0.416$ (crosses); and Cl~0016$+$16 at
$z=0.55$ (squares).  The solid line is a least squares bisector fit to
the cluster data.\label{kormendy2}}
\end{figure}

In Fig.\ \ref{kormendy2}, we compare the LBDS radio galaxies with
cluster ellipticals at $z\simeq 0.4$.  This sample consists of all the
early-type galaxies from \citet{Ziegler99}, drawn from five clusters
observed with WFPC2 --- Abell 370 ($z=0.375$), Cl~1447$+$26
($z=0.389$), Cl~0939$+$47 ($z=0.407$), Cl~0303$+$17 ($z=0.416$) and
Cl~0016$+$16 ($z=0.55$).  \citet{Ziegler99} corrected their data to
zero-redshift $B$-band, and compared the differences in average
surface brightness ($\Delta M$) between the $z\simeq 0.4$ clusters and
the Coma cluster (they used two different samples of Coma data and two
different measures of average surface brightness).  \citet{Ziegler99}
demonstrated that the slope of the Kormendy relation for each of these
clusters was consistent with that of Coma, and that the difference in
surface brightnesses $\Delta M$ was due to passive evolution of the
stellar populations, possibly with some low-level ongoing star
formation.  In Fig.\ \ref{kormendy2} we plot their fully corrected
data for our cosmology, using the `ComaSBD/$\Delta M_{\rm free}$'
surface brightness corrections from their table~4 (taking their
alternative $\Delta M$ values does not change our results).  Using a
least-squares bisector fit \citep{Isobe90} we find the best-fitting
Kormendy relation for the combined sample to be $\left<\mu_e\right> =
(2.8\pm0.2) \log r_e + (19.4\pm0.1)$.

For each of the two LBDS radio galaxies, we have measured the mean
surface brightness within $r_e$ from the F110W data
(Table~\ref{sbdata}) and corrected these $\left<\mu_e\right>$ values
to zero redshift.  Specifically, we applied corrections of: (i)
$+1.8$~mag K-correction from observed F110W to restframe $B$; (ii)
$+1.5$~mag of passive evolution from 3~Gyr ($z=1.5$) to 13~Gyr
($z=0$); (iii) $-0.7$~mag AB to Vega systems; and (iv) $-4.0$~mag due
to cosmological surface brightness dimming.  The combined error in
these corrections is $\pm0.3$~mag, noting that most of the uncertainty
in the models occurs at early ages and thus the error is no larger
than that in the previous section.  53W069 (solid square) and 53W091
(solid circle) are overplotted on the cluster Kormendy relation in
Fig.~\ref{kormendy2}.  The LBDS galaxies lie on the Kormendy relation,
entirely consistent with the scatter in the relation for cluster
ellipticals.

Our investigations of the Kormendy relation indicate that 53W069 \&
53W091 are passively-evolving, dynamically-relaxed ellipticals at
$z\simeq 1.5$.  However, the results of \citet{Ziegler99} show that
early-type galaxies with weak disk components and low-level star
formation, do not differ significantly from ellipticals on the
Kormendy relation.  Therefore the results of this section alone do not
exclude the possibility of some star formation in the LBDS galaxies;
indeed the disk component identified in 53W091 (section 3) may be the
site of recent low-level star formation activity.

\section{Scaling relations and masses}

We have shown above that the two $z\simeq1.5$ LBDS radio galaxies
53W069 \& 53W091 are significantly smaller than the average 3CR galaxy
at $z\simeq0.8$ (section 4.1).  In this section, we consider whether
their smaller size could also explain their lower radio luminosity.
We recall how both size and radio power scale with the mass of the
galaxy and its central black hole, and then estimate of the masses of
these two radio galaxies.

The correlation between the central black hole mass ($M_{\rm bh}$) and
the spheroid mass ($M_{\rm sph}$) of galaxies is now well-established
\citep{Magorrian98}:
\begin{equation}
M_{\rm bh} \propto M_{\rm sph}
\end{equation}
For spirals, $M_{\rm sph}$ is the mass of the bulge (spheroid)
component (i.e.\ excluding the disk), for ellipticals it is the total
(virial) mass of the galaxy.  Further investigations have also
revealed correlations between $M_{\rm bh}$ and spheroid luminosity
($L$) and between $M_{\rm bh}$ and stellar velocity dispersion
\citep{Kormendy01}.  In the absence of velocity dispersion data we
adopt the well-known relation between galaxy mass and luminosity,
using the mass-to-light ratio for elliptical galaxies from
\citet*{Jorgensen96}:
\begin{equation}
M_{\rm sph} \propto L^{1.3}
\end{equation}
For a de Vaucouleurs $r^{1/4}$ law, the integrated galaxy luminosity
is given by $L \propto I_e r_e^2$ where the intensity at the effective
radius ($I_e$) is trivially related to the effective surface
brightness $\mu_e \propto -2.5\log I_e$.  We adopt a Kormendy relation
with a slope of 3, $\mu_e \propto 3\log r_e$, this being consistent
with the 3CR results of \citet{McLure00}, the cluster relation found
in section 4.2 above and the $z\simeq 0.2$ AGN sample of
\citet{Dunlop02}.  Combining these equations thus gives a scaling
relation between the effective radius and the spheroid mass of
elliptical galaxies:
\begin{equation}
r_e \propto M_{\rm sph}^{1.0}
\end{equation}
The error in the index is due to errors in the $M_{\rm sph}$--$L$
correlation (e.g.\ \citealt{Kormendy01} find $M_{\rm bh} \propto
L^{1.1}$, c.f.\ equations 1 \& 2) and the slope of the Kormendy
relation ($\pm0.2$, see values in section 4).  The combination of
these errors leads to an error in the power-law index of no more than
$\pm0.2$.

Whatever the physical mechanism that drives the radio output of AGN,
and indeed non-active spheroids, a number of authors have found a
power-law dependence between radio power ($P$) and black hole mass:
\begin{equation}
P \propto M_{\rm bh}^\gamma \propto M_{\rm sph}^\gamma
\end{equation}
where we have used the Magorrian relation (equation 1) in the second
step.  \citet*{Franceschini98} find a best-fitting value of
$\gamma=2.7\pm0.3$ for a sample of nearby galaxies, whose black hole
masses are based directly on stellar dynamics.  They note that
$\gamma\simeq2$ is a generic prediction of any emission process that
depends on the emitting area around a black hole, and in particular
$\gamma=2.2$ is the prediction of advection-dominated accretion flows
\citep[see e.g.,][]{Narayan95}.  \citet{Lacy01} quote a value of
$\gamma=1.4\pm0.2$ for a sample of radio-loud quasars, although
inspection of their Fig.~2 suggests $\gamma\simeq2$--2.5 is also
consistent with their data.  \citet{Dunlop02} demonstrated that with
$\gamma=2.5$, equation 4 is consistent with both (i) normal galaxies
and radio-quiet quasars, and (ii) radio galaxies and radio-loud
quasars, but with a normalization that differs between the two classes
of radio power.

If equations 3 \& 4 are combined, we get a scaling relation between
the radio power and effective radius of radio-loud ellipticals:
\begin{equation}
P \propto r_e^{2.7}
\end{equation}
where we have used $\gamma=2.7$ from \citet{Franceschini98}.  Given
the uncertainties in $\gamma$ this can only be considered an
approximate relation, but it is still interesting to compare the two
LBDS galaxies with the more powerful 3CR sources at $z\simeq0.8$.  We
converted the observed 8.4~GHz flux densities of the ten 3CR sources
\citep*{Best97} to emitted 1.4~GHz luminosites using a mean spectral
index of $\alpha=0.8$ (where $S_\nu\propto\nu^{-\alpha}$).  Their mean
power is then $P_{\rm 3CR}=9\times10^{27}$~W~Hz$^{-1}$ in our
cosmology.  We similarly calculated the 1.4~GHz luminosites of 53W069
\& 53W091 (with spectral indices of 0.9 \& 1.1 respectively;
\citealt{Waddington00}; \citealt{Spinrad97}), giving
$P_{LBDS}\simeq5\times10^{26}$~W~Hz$^{-1}$.  Given these radio powers,
and recalling that the mean effective radius of the 3CR sources is
12~kpc (section 4.1), the scaling relation (equation 4) would predict
the sizes of the LBDS galaxies to be 4~kpc.  This is in excellent
agreement with their actual effective radii of 3~kpc.  The smaller
size and lower radio luminosity of these galaxies compared with 3CR is
consistent with them being less-massive galaxies with proportionately
less-massive black holes; their smaller size cannot be taken as
evidence of evolution in the size of the largest ellipticals.

We estimated the virial masses of 53W069 \& 53W091 as follows.
Although we do not know the velocity dispersions, we can estimate them
from the zero-redshift Fundamental Plane relation \citep{Jorgensen96}.
Using the effective radius and present-day $B$-band surface brightness
of each galaxy derived in section~4 ($\left<\mu\right>_e=20.4$ for
53W069 \& 20.0 for 53W091), we predict a velocity dispersion of
$\sigma\simeq150\pm45$~km~s$^{-1}$ for the LBDS galaxies.  The
present-day absolute $B$-band magnitudes (Vega system) were similarly
calculated from the observed F110W magnitudes and the corrections of
section 4.  We find $M_B=-20.2$ for 53W069 and $-20.3$ for 53W091,
corresponding to luminosities of $L_B = 0.9 L_B^*$ and $1.0 L_B^*$
respectively \citep{Madgwick02}.  Using the Faber-Jackson relation
recently measured for Coma and three $z\simeq0.4$ clusters
\citep{Ziegler97}, we estimate a velocity dispersion of $\sigma\sim
180$~km~s$^{-1}$ for the LBDS galaxies, consistent with the more
accurate determination from the zero-redshift Fundamental Plane.  The
virial mass, $M_{\rm vir}\sim 5\sigma^2 r_e / G$, is inferred to be
$M_{\rm vir}\sim (0.4$--$1.2) \times 10^{11}$~M$_{\sun}$ for each of
the two galaxies.  From equation (1), the mass of their central black
holes is inferred to be $(1$--$3) \times 10^8$~M$_{\sun}$, taking the
constant of proportionality to be 0.0025 \citep{McLure01,Dunlop02}.

\section{Conclusions}

We have used \hst\ observations of the two millijansky radio galaxies
LBDS 53W069 and 53W091 to investigate their optical morphologies at
$z\simeq 1.5$.  In both F814W (restframe $U$-band) and F110W
(restframe $B$-band) 53W069 is best described by an elliptical (de
Vaucouleurs) model, of effective radius $0\farcs3$ or 3~kpc.  The
$U-B$ colour gradient indicates that the galaxy formed in a rapid
burst of star formation at high redshift ($z>5$) and has a high
metallicity ($3Z_{\sun}$) stellar population.  In F110W, 53W091 is
similarly modelled as an elliptical galaxy of effective radius
$0\farcs3$ (3~kpc).  At shorter wavelengths (F814W), 53W091 is more
extended, consistent with there being a faint blue disk contributing
$\sim20$\% of the flux within $r_e$.  The $U-B$ colour of the galaxy
within a radius of 0\farcs4 indicates a high metallicity
($3Z_{\sun}$), high-redshift ($z>5$) burst of star formation, whereas
the metallicity falls to solar at larger radii.

Assuming passive evolution of their stellar populations, these two
elliptical galaxies lie on the $B$-band Kormendy surface
brightness--effective radius relations of both 3CR radio galaxies and
cluster ellipticals.  Their sizes and radio luminosities are
consistent with scaling relations applied to the 3CR radio galaxies,
in which both radio power and effective radius scale with galaxy mass.

Our analyses of the restframe UV spectra of these two galaxies
demonstrated that they were passively-evolving ellipticals, with ages
$\ga 3$~Gyr at $z\simeq 1.5$
\citep{Dunlop96,Spinrad97,Dunlop99b,Nolan02}.  The age of the universe
at this redshift is 4.5~Gyr in the flat, lambda-dominated cosmology
that we assume here, requiring a formation redshift of $z_f\ga 5$.  In
\citet{Peacock98} we compared the primordial density fluctuation
spectrum required by the existence of these galaxies with that
inferred from Lyman-$\alpha$ absorbers and Lyman-break galaxies.
Those results indicated a similarly high redshift of gravitational
collapse, $z_c\simeq 6$--8.

Here we have demonstrated that the morphologies and internal colours
of these galaxies similarly show them to be ellipticals whose last
major episode of star formation was at very high redshift.  These
results imply that the last major merger was also some while in the
past, since such an event would produce significant star formation and
an irregular morphology, for which we see no strong evidence at the
epoch at which we see them.  Although it may be that the faint blue
disk component in 53W091 is the remains of the galaxy's last merger.
In this context, we note that 53W091 has several companions with
similar colours, one of which has been spectroscopically confirmed to
have the same redshift of 1.55 \citep{Spinrad97,Bunker00}.  It is
possible, even likely, that one or more of these companions will merge
with 53W091 in the future.  In contrast, 53W069 has no such
companions.

We began the paper by asking how ellipticals formed -- via monolithic
collapse or hierarchical mergers?  We end by concluding that we cannot
be certain, even in the particular case of these two LBDS radio
galaxies.  The most likely interpretation of the data is that they
have not evolved significantly since high redshifts ($z\sim 5$), save
for passive stellar evolution.  However, their formation could still
have been via mergers at early times, and at least in the case of
53W091, it seems probably that further mergers lie ahead.  What is
clear, is that these radio galaxies are quite ordinary ellipticals,
and if they continue to evolve passively until the present day they
will become typical $L^*$ elliptical galaxies.

\section*{Acknowledgments}

This work was based on observations with the NASA/ESA {\it Hubble
Space Telescope\/} obtained at the Space Telescope Science Institute,
which is operated by the Association of Universities for Research in
Astronomy, Inc., under NASA contract NAS5-26555.  Support was provided
by NASA grant GO-7280.*.96A from STScI under NASA contract NAS5-26555.
IW acknowledges the support of the Leverhulme Trust.  JSD acknowledges
the enhanced research time afforded by the award of a PPARC Senior
Fellowship.  RJM acknowledges the award of a PPARC Postdoctoral
Fellowship.  The work of DS was carried out at the Jet Propulsion
Laboratory, California Institute of Technology, under a contract with
NASA.

\bsp

\label{lastpage}

\end{document}